\documentclass[aps,prl,twocolumn,superscriptaddress,longbibliography,floatfix]{revtex4-2}
\usepackage{graphicx,bm,amssymb,amsmath,dcolumn,hyperref}
\usepackage{multirow}
\usepackage{enumerate}
\usepackage{enumitem}
\usepackage{color}
\usepackage{subfigure}  
\usepackage{epstopdf}
\usepackage{bbm}
\usepackage{graphicx}
\usepackage{threeparttable}
\usepackage{amsthm}
\usepackage{mathtools}
\usepackage{color}
\usepackage{tikz}
\usepackage{float}
\usepackage{empheq}     
\usepackage{booktabs}

\draft 

\begin{document}

\preprint{}

\title{
Is the atmospheric river operating at a self-organized criticality state?
} 



\author{Shang Wang}
\affiliation{School of Systems Science/Institute of Nonequilibrium Systems, Beijing Normal University, Beijing 100875, China}

\author{Jun Meng}
\affiliation{Key Laboratory of Earth System Numerical Modeling and Application, Institute of Atmospheric Physics, Chinese Academy of Sciences, Beijing 100029, China}

\author{Sheng Fang}
\affiliation{School of Systems Science/Institute of Nonequilibrium Systems, Beijing Normal University, Beijing 100875, China}

\author{Teng Liu}
\affiliation{School of Systems Science/Institute of Nonequilibrium Systems, Beijing Normal University, Beijing 100875, China}
\affiliation{Earth System Modelling, School of Engineering and Design, Technical University of Munich, Munich, 85748, Germany}
\affiliation{Potsdam Institute for Climate Impact Research, Potsdam 14412, Germany}

\author{Kim Christensen}
\affiliation{Blackett Laboratory and Centre for Complexity Science, Imperial College London, South Kensington Campus, London SW7 2AZ, United Kingdom}

\author{J\"urgen Kurths}
\affiliation{Potsdam Institute for Climate Impact Research, Potsdam 14412, Germany}
\affiliation{Department of Physics, Humboldt University, Berlin 10099, Germany}

\author{Jingfang Fan}
\email{jingfang@bnu.edu.cn}
\affiliation{School of Systems Science/Institute of Nonequilibrium Systems, Beijing Normal University, Beijing 100875, China}
\affiliation{Potsdam Institute for Climate Impact Research, Potsdam 14412, Germany}


\date{\today}

\begin{abstract}
Atmospheric rivers (ARs) are essential components of the global hydrological cycle, with profound implications for water resources, extreme weather events, and climate dynamics. Yet, the statistical organization and underlying physical mechanisms of AR intensity and evolution remain poorly understood. Here we apply methods from statistical physics to analyze the full life cycle of ARs and identify universal signatures of self-organized criticality (SOC). We demonstrate that AR morphology exhibits nontrivial fractal geometry, while AR event sizes—quantified via integrated water vapor transport—follow robust power-law distributions, displaying finite-size scaling. These scaling behaviors persist under warming scenarios, suggesting that ARs operate near a critical state as emergent, self-regulating systems. Concurrently, we observe a systematic poleward migration and intensification of ARs, linked to thermodynamic amplification and dynamical reorganization.  Our findings establish a statistical physics framework for ARs, linking critical phenomena to the spatiotemporal structure of extreme events in a warming climate.
\end{abstract}

\maketitle 

Many complex natural systems exhibit self-organized criticality (SOC), a state where they naturally evolve toward a critical point through multiscale interactions, giving rise to emergent, scale-invariant behaviors without requiring external fine-tuning~\cite{bak_self-organized_1987}.
SOC has been widely documented in earthquake dynamics, sandpile avalanches, forest fires, and $1/f$ noise in condensed matter systems~\cite{bak_self-organized_1988,sornette_self-organized_1989,jensen_self-organized_1998,christensen2005complexity}. These systems are characterized by power-law distributions in event sizes, durations, and energy dissipation, reflecting their underlying near-critical dynamics~\cite{peters_complexity_2001,sanchez_waitingtime_2002}.
In Earth system science, SOC has been proposed as a governing mechanism for phenomena such as seismicity~\cite{sornette_self-organized_1989}, turbulent cascades~\cite{frisch1995turbulence,smyth_self-organized_2019}, and climate fluctuations~\cite{lovejoy2013weather}, with recent work emphasizing statistical physics as a unifying approach to complex Earth system~\cite{fan_statistical_2021}. Although critical phenomena like SOC have been proposed for rain (precipitation) in highly cited papers~\cite{peters_complexity_2001,Peters2002PRE,peters_critical_2006}, research on its application to atmospheric processes remains relatively limited.
Many geophysical systems are governed by energy dissipation, non-linear transport and scale invariance. This leads to a fundamental question: Do atmospheric rivers, as the main transporters of moisture, exhibit SOC?

\textit{Atmospheric rivers} (ARs) are defined as long, narrow transient water vapor corridors that play an important role in the Earth's hydrological cycle, accounting for over 90\% of poleward atmospheric moisture transport~\cite{zhu_proposed_1998,guan_detection_2015,nash_role_2018}.
They significantly influence mid- and low-latitude regions, such as Europe~\cite{lavers_contribution_2015,scholz_atmospheric_2024}, North America~\cite{lavers_contribution_2015,vallejo-bernal_spatio-temporal_2022}, East Asia~\cite{pan_east_2020}, and India~\cite{mahto_atmospheric_2023}. They provide essential precipitation that supports agriculture and economic development. 
However, ARs are also a leading source of extreme rainfall, triggering floods, landslides, and debris flows that pose significant risks to lifes and infrastructure. 
Moreover, ARs act as long-range aerosol transport vectors, such as in the case of Saharan dust reaching Europe, which has implications for air quality, public health, and broader ecological systems~\cite{francis_atmospheric_2022}.
Recent research has also highlighted the growing influence of ARs at high latitudes. AR incursions into the Arctic and Antarctic contribute to rapid ice sheet and glacier melting, while also supporting seasonal snowpack accumulation in alpine regions~\cite{zhang_more_2023,liang_role_2023,francis_crucial_2020,wille_intense_2022,guan_extreme_2010}.

\textit{Variabilities and uncertainties of ARs.} ARs serve as major conveyor belts linking oceanic evaporation to continental precipitation and are shaped by a combination of thermodynamic and dynamic processes. 

Thermodynamically, AR behavior is governed by the Clausius-Clapeyron relation, which predicts an exponential increase in saturation vapor content $q^*$ with temperature $T$ as~\cite{payne_responses_2020}, 
\begin{equation}
    \frac{dq^*}{dT} = \alpha (T) q^*,
    \label{eq:eq1}
\end{equation}
where $\alpha (T) = \frac{L}{R_v T^2}$ is the Clausius-Clapeyron scaling factor, $L$ is the latent heat of vaporization and $R_v$ is the gas constant of water vapor. This relationship highlights the thermodynamic amplification of water vapor transport as temperatures rise.

Dynamically, AR evolution is modulated by large-scale circulation patterns, which are more uncertain under climate change. The quasi-geostrophic potential vorticity equation captures key aspects of AR-modulated vorticity dynamics~\cite{vallis_atmospheric_2017},
\begin{equation}
\frac{\partial \zeta}{\partial t} + \vec{v} \cdot \nabla \zeta + \beta v + f \frac{\partial w}{\partial p} = - \nabla \cdot \left( \frac{\vec{J}_m}{\rho} \right),
\label{eq:dynamic_response}
\end{equation}
where \( \zeta \) represents the relative vorticity, \( \vec{v} = (u, v) \) is the horizontal wind vector (with zonal \( u \) and meridional \( v \) components), \( \beta = \frac{\partial f}{\partial \phi} \) represents the latitudinal variation of the Coriolis parameter \( f \), \( w \) is the vertical velocity in pressure coordinates, \( p \) is pressure, \( \vec{J}_m \) is the moisture flux vector, and \( \rho \) is the atmospheric density. Unlike the relatively well-understood thermodynamic processes, the dynamic response of atmospheric circulation to warming is less certain, as it is closely linked to complex features like vortex-driven jet streams~\cite{payne_responses_2020,ma_poleward_2020}. 


While numerous studies have explored AR trends using reanalysis data and climate models~\cite{payne_responses_2020,wang_role_2024,scholz_atmospheric_2024}, the results remain inconsistent due to methodological differences in AR detection~\cite{guan_detection_2015,xu_image-processing-based_2020,xu_ipart_2020}. More critically, there is a lack of a physical framework that captures the emergent, nonlinear behavior of ARs. In this Letter, we address this gap by applying a statistical physics perspective to the full life cycle of ARs. This approach provides a dynamical, macroscopic view of AR variability, while minimizing sensitivity to detection thresholds and dataset inconsistencies.

ARs are characterized by their pronounced moisture transport, defined as the rate of mass (water vapor) transfer across a unit area, which is quantitatively captured by the column-integrated water vapor transport (\textit{IVT}, \({\rm kg} \cdot {\rm m}^{-1} \cdot {\rm s}^{-1}\)), and mathematically expressed as follows:
\begin{equation}
    IVT = \frac{1}{g} \sqrt{\left( \int_{p_{b}}^{p_{u}} q u \, dp \right)^2 + \left( \int_{p_{b}}^{p_{u}} q v \, dp \right)^2},
    \label{eq:ivt}
\end{equation}
where \( g \) is the gravitational acceleration, \( q \) is the specific humidity, \( p_{b} \) and \( p_{u} \) represent the surface pressure and an upper-atmospheric reference pressure, respectively. The moisture flux $\vec{J}_m$, central to AR dynamics, is given by
\begin{equation}
\vec{J}_m = \rho q \vec{v}, 
\label{eq:JM} 
\end{equation} 
linking the atmospheric circulation to water vapor transport in Eq.~(\ref{eq:dynamic_response}).

\begin{figure}[htbp]
      \centering
      \includegraphics[width=1\linewidth]{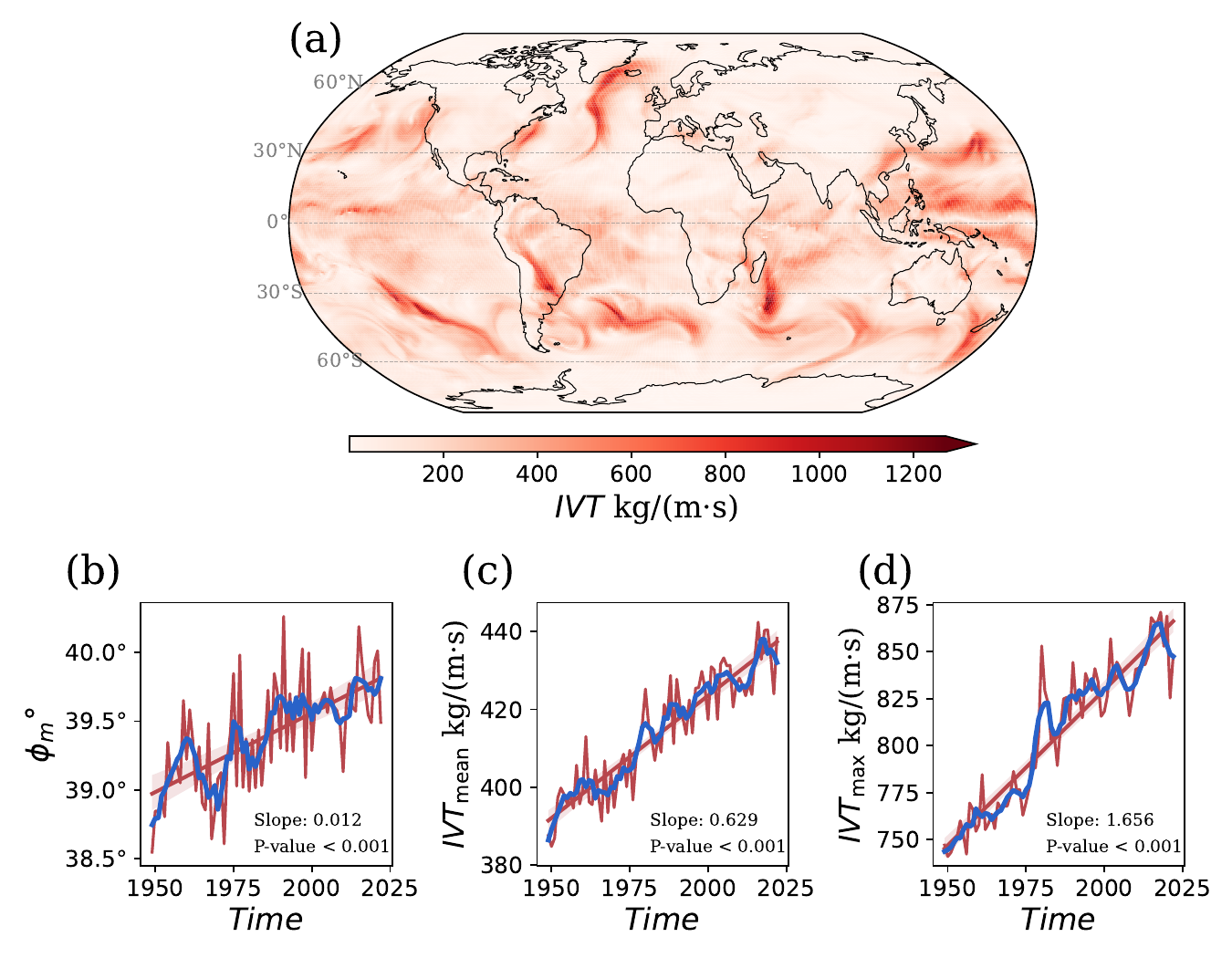}
      \caption{
        {Climatic variability and trends in AR events over last seven decades.}
        (a) Conceptual diagram of ARs based on a real \textit{IVT} distribution captured at 6:00 AM on January 9, 2019.
        (b) Annual variation in the latitudinal position $\phi_{m}$ of AR events at their peak $IVT_{\rm{max}}$ intensity during their lifetime. 
        (c-d) Annual mean values and trends in $IVT_{\rm{mean}}$ and $IVT_{\rm{max}}$ intensity throughout the life cycle of AR events.
        The red solid line represents the linear regression fit, with the shaded area indicating the 95\% confidence interval. The solid blue line depicts a five-year moving average.}
        \label{fig1}
    \end{figure}

We identify ARs using the IPART algorithm~\cite{xu_ipart_2020} on 6-hourly high spatiotemporal resolution \textit{IVT} data derived from the ERA5 dataset (1949–2022)~\cite{hersbach_era5_2020}, as defined in Eq.~(\ref{eq:ivt}). Figure~\ref{fig1}(a) illustrates a representative AR structure, from January 9, 2019, at 06:00 UTC, revealing its characteristic narrow geometry and high transport intensity.
Methodological details, ARs detection method, dataset information, and robustness analyses are provided in the Supplemental Information~\cite{SM}.

\textit{Climate-driven shifts in AR characteristics.} 
We define the maximum integrated vapor transport during each AR's lifetime as $IVT_{\rm{max}}$, with the corresponding latitude denoted by $\phi_{m}$, and the temporal average as $IVT_{\rm{mean}}$. 
Figure~\ref{fig1}(b) demonstrates a pronounced poleward migration of $\phi_{m}$ (expressed in absolute values), advancing by $0.12^\circ$ per decade, with a particularly strong signal in the Southern Hemisphere (see Fig. S1).
This shift reflects a systematic reorganization of the global atmospheric circulation.

The observed migration is driven by coupled thermodynamic and dynamic processes~\cite{li_global_2024}. A key contributor is the poleward expansion of the Hadley Cell under warming, which displaces the subtropical high-pressure belt and consequently shifts AR pathways~\cite{fu_enhanced_2006,lu_expansion_2007,fan_climate_2018}. Concurrently, intensifying extratropical cyclones exhibit a latitudinal shift toward inland and polar regions~\cite{li_recent_2023,kossin_poleward_2014,emanuel_increasing_2005,lin_poleward_2023,studholme_concurrent_2018,feng_poleward_2021}. These changes are amplified by weakening of the polar vortex and expansion of polar low-pressure systems, which increase AR incursions into higher latitudes. AR evolution is also modulated by Rossby wave dynamics and midlatitude jet streams, both of which have become increasingly variable under global warming. Additionally, Arctic Amplification—characterized by rapid polar warming and sea ice loss—alters the strength and position of the polar jet~\cite{zhang_more_2023,liang_role_2023,screen_central_2010,stuecker_polar_2018}, reinforcing shifts in AR tracks. Together, these mechanisms point to a robust dynamical response of AR systems to anthropogenic climate forcing.

Figures~\ref{fig1}(c) and (d) show  a significant increase in both $IVT_{\rm{mean}}$ and $IVT_{\rm{max}}$ of ARs across recent decades. This trend is primarily thermodynamic: as described by the Clausius-Clapeyron relation, Eq.~(\ref{eq:eq1}), saturation vapor pressure increases by approximately 7\% per $^\circ C$~\cite{payne_responses_2020}. Elevated sea surface temperatures amplify evaporation, particularly in tropical and subtropical regions, supplying more moisture for AR development. 
El Niño events amplify AR intensity by generating warm sea surface temperature anomalies in the central and eastern tropical Pacific, which increase water vapor availability and AR frequency, particularly along the U.S. West Coast~\cite{mahto_atmospheric_2023,corringham_atmospheric_2019}. Beyond thermodynamic factors, anthropogenic influences such as urbanization and land-use change may alter regional circulation and vapor distribution, with potential impacts on AR behavior. Additionally, long-term shifts in mid-latitude atmospheric circulation~\cite{butler_steady_2010}, along with large-scale climate oscillations like the Pacific Decadal Oscillation and Atlantic Multidecadal Oscillation ~\cite{zhu_proposed_1998,lavers_nexus_2013,dettinger_fifty-two_2004}, contribute to the observed intensification and extended duration of AR events. 



\textit{Critical geometry of ARs.} The irregular geometry of AR structures prompts the question: do ARs exhibit fractal characteristics? To assess this, we compute the fractal dimension $D_f$ by examining the scaling relation between the AR area $A$ and the radius of gyration $R$, defined as,
   \begin{equation}
        R = \sqrt{ \frac{1}{N} \sum_{i=1}^N (r_i - \bar{r} )^2 },
        \label{eq:eq3}
    \end{equation}
where $N$ is the total number of points in the AR region, $r_i$ denotes their positions, and $\bar{r} = \frac{1}{N} \sum_{i=1}^N r_i$ is the centroid.

\begin{figure}[htbp]
    \centering
    \includegraphics[width=1\linewidth]{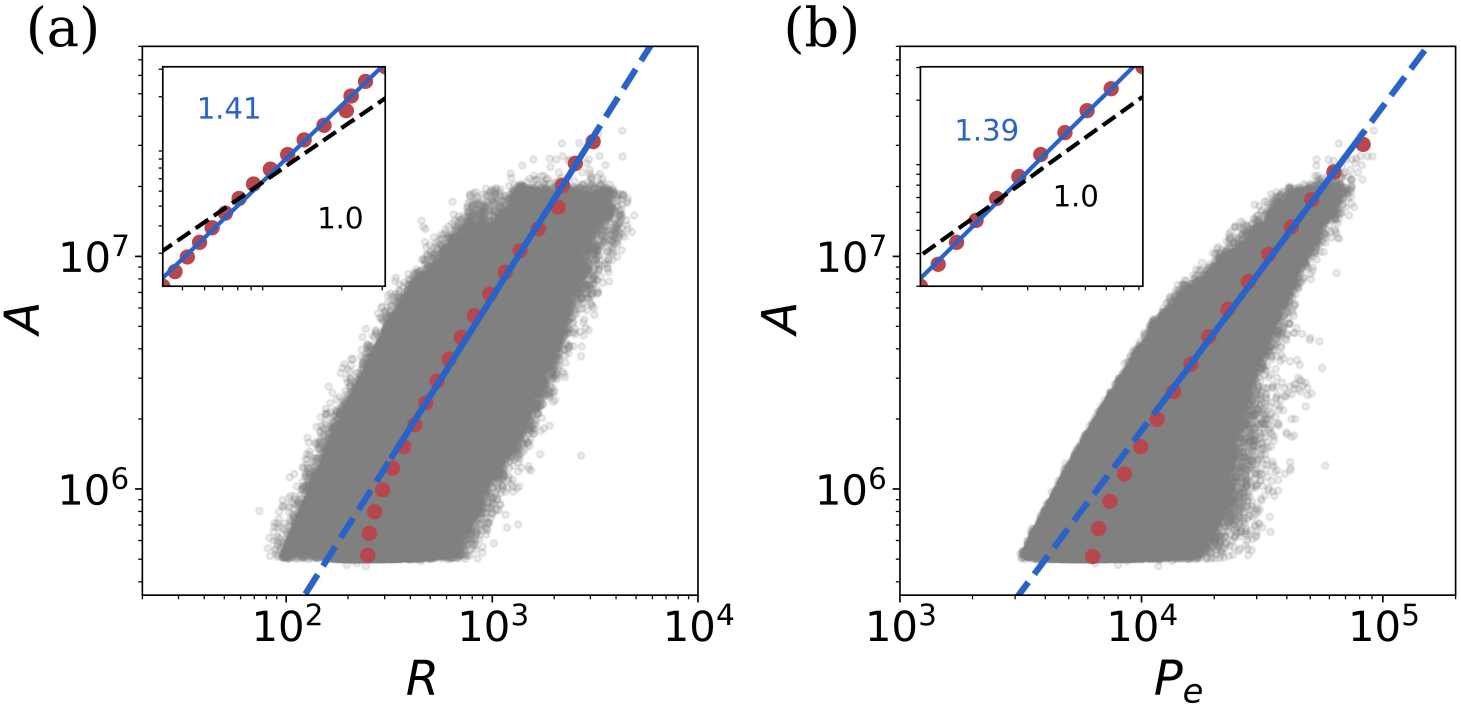}
    \caption{
    {Scaling relationships between the area ($\it{A}$) of ARs and their radius of gyration ($\it{R}$) and perimeter ($\it{P_e}$).}
    (a) Power law relationship between $\it{R}$ and $\it{A}$ with a power-law exponent $D_f = 1.41 \pm 0.03$. 
    (b) Power law relationship between $\it{P_e}$ and $\it{A}$ with a power-law exponent  $D_p = 1.39 \pm 0.03$.
    Shaded points in the background represent the original data distribution, solid red points show the binned logarithmic averages, and dashed blue lines indicate fitted slopes corresponding to $D_f$ and $D_p$. The inset zooms into the power-law region, with the black dashed line serving as a reference slope of 1.0.}
    \label{fig2}
\end{figure}

Computing $A$ and $R$ for all AR events over the past 74 years (gray dots in Fig.~\ref{fig2}(a), including 1,403,699 data records corresponding to 194,774 distinct AR events), we observe robust a power-law scaling. The ensemble-averaged relation (red dots) yields a fitted exponent $D_f = 1.41(3)$, significantly deviating from the trivial Euclidean scaling $D_f=1$, as highlighted by the black dashed guideline in the inset. We also examine the perimeter-area relation and extract a fractal-like exponent $D_p = 1.39(3)$, as shown in Fig.~\ref{fig2}(b). The consistency of both exponents across events and detection methods supports the presence of this nontrivial spatial organization. These results suggest that ARs form scale-invariant structures, analogous to critical clusters in percolation theory~\cite{stauffer_introduction_2018,ben2000diffusion}.

    \begin{figure*}[htbp]
      \centering
      \includegraphics[width=1.0\linewidth]{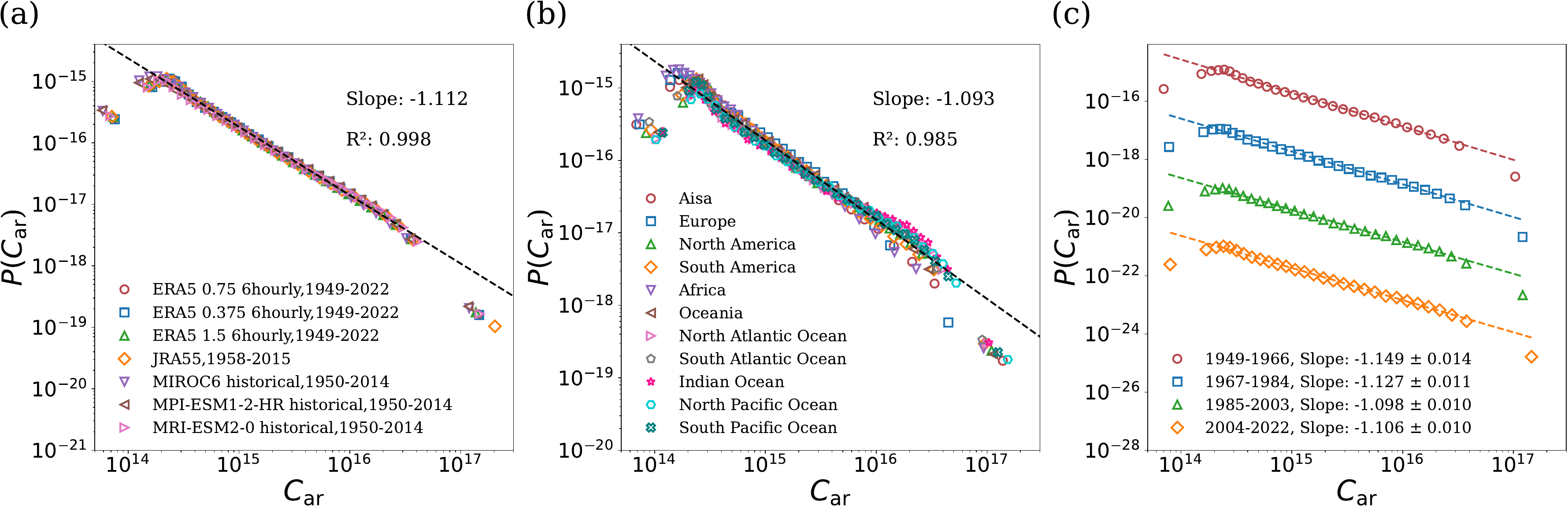}
      \caption{ Power-law distribution of total water vapor transport capacity $\it{C}_{\rm{ar}}$.
        (a) Probability density function $\it{P(C_{\rm{ar}})}$ of $\it{C}_{\rm{ar}}$ for AR events presented in log-log coordinates based on different spatiotemporal datasets.
        (b) $\it{P(C_{\rm{ar}})}$ of $\it{C}_{\rm{ar}}$ for AR events across various geographical regions during 1949–2022, using ERA5 data. Detailed information on the geographical divisions is provided in SI~\cite{SM}.
        (c) $\it{P(C_{\rm{ar}})}$ of $\it{C}_{\rm{ar}}$ for AR events over different time periods, highlighting temporal variability with an artificial difference of $10^{-2}$.
        }
        \label{fig3}
    \end{figure*}

\textit{Universal scaling of AR dissipation.} A hallmark of complex systems is the emergence of scale-free distributions in event magnitudes. To quantify AR event size, we define a macroscopic indicator of the total water vapor transport capacity, \( C_{\rm{ar}} \), as
\begin{equation}
    C_{\rm{ar}} = \sum_t \sum_i A_{i,t} \cdot IVT_{i,t},
    \label{eq:eq4}
\end{equation}
where \( A_{i,t} \) is the area associated with grid point \( i \) at time \( t \) (adjusted for latitude via \( \cos \phi \)), and \( IVT_{i,t} \) is the corresponding IVT intensity. AR tracks are sampled at 6-hour intervals with no imposed duration threshold, ensuring a comprehensive accounting of event sizes, including short-lived filamentary structures.

We find that \( C_{\rm{ar}} \) follows a robust power-law distribution (Fig.~\ref{fig3}),
\begin{equation}
    P(C_{\rm{ar}}) \propto C_{\rm{ar}}^{-\tau} \, g\left( \frac{C_{\rm{ar}}}{C^{*}_{\rm{ar}}} \right),
    \label{eq:eq5}
\end{equation}
where \( \tau \approx 1.1 \) is the scaling exponent, \( g(x) \) is a rapidly decaying function for \( x \gg 1 \), and \( C^{*}_{\rm{ar}} \) denotes the effective upper cutoff of event size due to finite-size effects. This formulation captures the scale-invariant regime for intermediate event sizes and the exponential suppression of rare, basin-limited ARs. The exponent remains consistent across spatial and temporal partitions [Figs.~\ref{fig3}(b)--(c)], indicating the universality of the scaling law across the climate system. This universality indicates that ARs exhibit scale-invariant dynamics characteristic of self-organized critical systems~\cite{sornette_self-organized_1989}. Furthermore, the distribution of \( IVT_{\rm{max}} \) is well-described by a Gumbel form (Fig. S6), reinforcing the link between AR statistics and classical extreme value theory.

The absence of a characteristic scale in \( C_{\rm{ar}} \), together with a finite-size cutoff at large values, mirrors the behavior observed in percolation clusters and typical critical phenomena~\cite{stauffer_introduction_2018,ben2000diffusion}. In addition, AR duration and cumulative influence area both exhibit power-law distributions (Figs. S8--S9), reinforcing the interpretation that AR dynamics are governed by emergent, self-regulating processes consistent with SOC.

\textit{SOC and physical mechanisms of AR dynamics.} The emergence and dissipation of ARs reflect a self-organized regulatory mechanism within the coupled ocean–atmosphere system, governed by both thermodynamic constraints and large-scale circulation dynamics. Moisture accumulation is driven by surface evaporation and horizontal vapor convergence, enhanced by Clausius–Clapeyron scaling [Eq.(\ref{eq:eq1})], which predicts an exponential increase in saturation vapor pressure with temperature. This buildup is modulated by atmospheric transport processes described by the quasi-geostrophic vorticity balance [Eq.(\ref{eq:dynamic_response})], where wind shear, baroclinicity, and vertical motion govern moisture redistribution.

AR events emerge when this gradually accumulated moisture is released through rapid condensation and precipitation, often triggered by frontal lifting or orographic ascent. Once initiated, the evolution of an AR follows a ``trigger–propagation–dissipation" sequence, where moisture is transported and redistributed over synoptic scales before decaying through precipitation and dispersal. This sequence parallels classical SOC models~\cite{bak_self-organized_1987}, in which a system accumulates energy until local thresholds are exceeded, triggering a cascade of activity that dissipates accumulated stress.

In an AR system, the feedback between moisture transport, latent heat release, and circulation reorganizations provides a natural self-regulating loop. Local perturbations—such as transient low-pressure systems or jet stream undulations—can propagate spatially, modulating IVT over large areas. These perturbations are neither purely stochastic nor externally imposed, but arise from internal dynamics near criticality, where the system exhibits maximal susceptibility to fluctuations~\cite{corral_scaling_2010}. The resulting scale-free distribution of AR event sizes and durations reflects a metastable system operating near a critical threshold, analogous to avalanching in sandpile models or energy bursts in magnetized plasmas.

\begin{figure*}[htbp]
      \centering
      \includegraphics[width=0.8\linewidth]{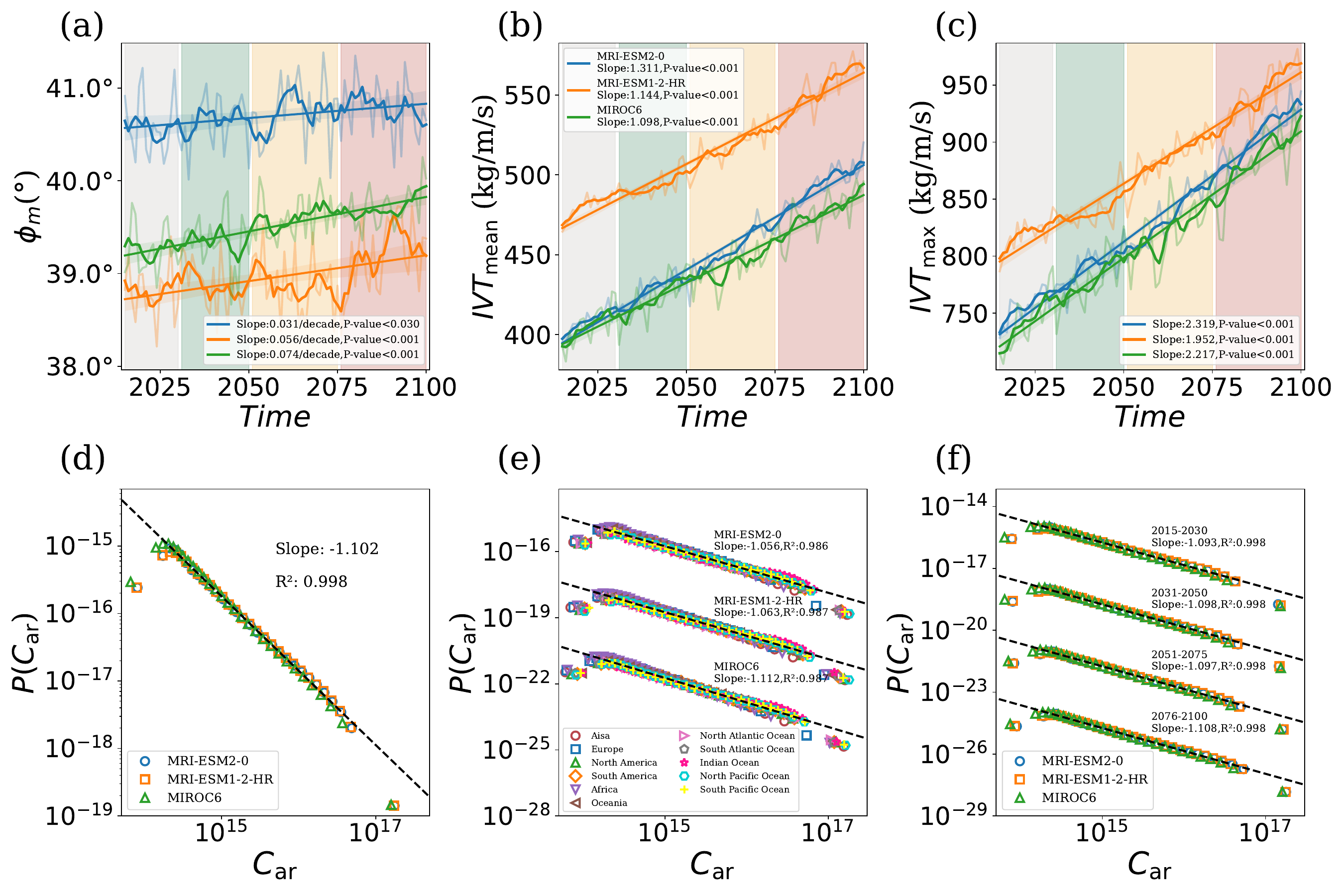}
      \caption{Response of AR climate variability to future warming scenarios. 
(a--c) Trends in three key AR metrics under CMIP6 SSP5-8.5 projections, analogous to Fig.~\ref{fig1}. Background shading indicates distinct future time periods. Light-colored lines (blue, green, orange) show raw data trends; solid lines with shaded envelopes denote linear fits with 95\% confidence intervals. Darker curves represent 5-year moving averages. 
(d--f) Probability density functions of \( C_{\rm{ar}} \), analogous to Fig.~\ref{fig3}, under different spatial and temporal partitions of the same warming scenario: (d) global across time, (e) regional across time, and (f) global across epochs. }
        \label{fig4}
    \end{figure*}

\textit{Response of ARs to future warming scenarios.} To evaluate how AR dynamics respond to long-term climate forcing, we analyze simulations from the Coupled Model Intercomparison Project Phase 6 (CMIP6)~\cite{gmd-9-1937-2016} under the SSP5-8.5 scenario—a high-emission pathway projecting substantial global warming through whole 21st century. Under this forcing, ARs exhibit a consistent poleward shift in the latitude of maximum IVT (\( IVT_{\rm{max}} \); Fig.~\ref{fig4}(a)), driven by weakened meridional temperature gradients and reorganization of large-scale circulation. This displacement extends AR influence into higher latitudes, increasing the frequency and intensity of extreme moisture transport events. Both \( IVT_{\rm{mean}} \) and \( IVT_{\rm{max}} \) increase significantly under warming (Figs.~\ref{fig4}(b)--(c)), reflecting thermodynamic amplification of atmospheric moisture. These trends raise the likelihood of high-impact precipitation events, with implications for hydrological extremes and regional vulnerability. Importantly, the total transport capacity \( C_{\rm{ar}} \) retains a robust power-law distribution across all models and future periods (Figs.~\ref{fig4}(d)--(f)), with finite-size scaling preserved. This invariance underscores the universality of AR scaling behavior and supports the interpretation of ARs as SOC systems under climate change. These findings suggest that despite thermodynamic and dynamic shifts, the statistical physics governing AR organization remains intact, highlighting the resilience of SOC behavior even in a warming world.

\textit{Discussion.} This study established that ARs operate near a state of SOC, as evidenced by robust nontrivial fractal geometry, power-law event size distributions, and finite-size scaling across spatiotemporal domains. These findings reveal ARs as emergent, self-organizing phenomena within the coupled ocean-atmosphere system, governed by universal scaling laws characteristic of critical systems. Our results demonstrate that global warming induces a systematic poleward migration of ARs, driven by weakened tropical-to-polar temperature gradients and shifts in atmospheric circulation. This migration, along with a significant increase in AR intensity, highlights the profound influence of climate change on AR dynamics.

The coexistence of SOC-driven universal scaling laws and climate variability represents a dialectical unity in atmospheric dynamics. While ARs exhibit a robust power-law statistics typical of critical systems, their spatiotemporal patterns remain sensitive to external forcings and regional climate variability. This interplay offers new insights into climate extremes: extreme AR events arise naturally from the internal variability of the SOC state, whereas climate change modulates their frequency and spatial reach. 

Physically,  the evolution of ARs follows a ``trigger–propagation–dissipation'' structure consistent with SOC dynamics. Initiation is driven by evaporation over warm SST anomalies, propagation occurs via synoptic-scale advection along jet streams, and dissipation results from latent heat release and precipitation. These stages form a dynamic cycle of moisture accumulation and release that parallels classical SOC cascades. This opens a pathway toward theoretical models of AR evolution, where moisture flux behaves analogously to energy in sandpile systems or percolation networks, and perturbations evolve through near-critical thresholds.

Future work may formalize this picture in a theoretical framework, e.g., a minimal stochastic threshold model governed by coupled thermodynamic forcing and atmospheric transport, to connect macroscopic scaling laws with microscale physics. Our findings bridge statistical physics and climate science, showing how universal principles manifest in Earth's complex climate system.

\section*{Acknowledgments}
We thank Dr. Guangzhi Xu for providing the IPART atmospheric river identification algorithm, and Dr. Shuyu Wang and Prof. Xiaohui Ma for their assistance with AR detection. We are grateful to Prof. Youjin Deng for valuable discussions on critical phenomena and self-organization in atmospheric dynamics. We also acknowledge the Atmospheric River Tracking Method Intercomparison Project (ARTMIP) for providing access to standardized AR detection datasets and tools. This work was supported by the National Natural Science Foundation of China (Grants No. 42450183, 12275020, 12135003, 12205025, 42461144209). J.F. acknowledges support from the Fundamental Research Funds for the Central Universities.



%
%

%


\bibliography{ref}
\bibliographystyle{apsrev4-2}




\end{document}